# PVC composite internal process of current-voltage time delay formation and conductivity levels lifetimes origin

D. V. Vlasov

Institute of General Physics

Abstract

In the previously described model of interpreting the numerous results in the switching states of the conduction in the films of PVC composites [1,2] was noted the need to involve hopping mechanism and percolation theory to explain the phenomenon of giant jumps conductivity, just as was done for the interpretation of such jumps for disordered semiconductor films. It was also shown that to explain the totality of spontaneous and stimulated transitions between states of the conduction in PVC composites requires the existence of an internal process that is able to produce a fixed observable current-voltage delay and switching electrical conductivity levels lifetime. In this paper one of the possible relaxation processes that lead to the emergence of the delay time, and the occurrence of spontaneous and stimulated transitions sample conductivity of nanocomposites PVC is presented. In particular, as a mechanism for determining time delay and lifetimes of the conduction states the reaction of dehydrochlorination is proposed, which should be reversible in case when HCl molecules separated from macromolecule physically can not "travel" far from "their place" in the macromolecule, as it just takes place within a local PVC free volume. In a frame of developed model all previously observed and called "anomalous" electrical phenomena receive a reasonable and predictable behavior.

## Введение

В изложенной авторами ранее [1,2] модели интерпретируются многочисленные результаты по переключению состояний проводимости, наблюдавшиеся экспериментально в полимерах и неупорядоченных полупроводниках (см. обзоры [3,4] и публикации [5,14]). В большинстве опубликованных работ речь идет об измерении тех или иных вольт-амперных характеристик, причем для интерпретации аномально высоких значений проводимости в тонких (толщиной, как правило, менее микрона) пленках привлекаются различные механизмы [3], начиная с инжектирования и формирование зон приэлектродного пространственного заряда до формирования наноразмерных металлических перемычек. Попытки построения моделей и понимания физических механизмов возникновения аномальных явлений в полимерных нанокомпозитах делались неоднократно с момента их обнаружения [5,6], причем предложенные модели отличаются большим разнообразием: от тривиальных пор или «дыр» в пленке образца до проявления уникальной сверхпроводимости в структуре металл-полимер-металл МПМ или левитации электронов [3]. Тем не менее, по признанию авторов наиболее полного на сегодняшний день обзора: **«Электроника тонких слоев широкозонных полимеров»** «… ситуация с понятными физическими механизмами аномальных явлений продолжает оставаться за рамками реального модельного физического описания даже на качественном уровне» [3]. В частности, авторы [3] неоднократно отмечают, что: «..**до сих пор вопрос об основных закономерностях электронного переключения и механизмах высокой проводимости полимерных тонкопленочных систем остается открытым**». Эти утверждения в полной мере можно отнести и к аномальным результатам скачков проводимости в относительно толстых пленках ПВХ композитов [7-13]. Тем не менее, в соответствии с [1,2] интерпретация переходов к аномально высокой проводимости в пленках полимеров или аморфных полупроводников с необходимостью включает два фундаментальных подхода. Первый, называемый в серии работ [7-13] «мягкий» обратимый пробой, т.е. резкое увеличение проводимости любого пленочного полимерного образца или неупорядоченного полупроводника при увеличении приложенного напряжения. Этот механизм связан со значительным увеличением вероятности туннелирования электрона из примесных центров, ловушек, дефектов структуры и т.д. при увеличении локального электрического поля Этот процесс давно и хорошо исследован и носит название эффекта Пула-Френкеля. В рамках этого механизма находит объяснение резкое падение сопротивления пленочных образцов связанное с нелинейным по электрическому полю ростом числа свободных носителей заряда в зоне проводимости и возникновении соответствующего тока, экспоненциально быстро увеличивающегося при увеличении приложенного напряжения. В определенном смысле этот механизм описывает начальное развитие лавинообразного процесса, который может при определенных условиях привести к пробою и необратимым разрушениям образца. Второй классический механизм скачков электропроводности связан с теорией протекания (порог перколяции), применимой скорее к композитным материалам, содержащим проводящие элементы и изолирующие (полимерные) промежутки (см. обзор [4] и подробную работу [14]). Этот механизм скачков проводимости реализуется и в произвольно слабых полях, и также часто привлекается для интерпретации скачкообразных изменений проводимости [7,10]. Следует отметить, что экспериментально наблюдаемые скачкообразные изменения проводимости полимерных материалов могут возникать не только при увеличении приложенного поля, но и при приложении одноосного давления [6], подсветке образца или изменении структуры электродов [3], а также ряда других внешних воздействий (описанных и систематизированных в [3]), что отчасти усложняет их интерпретацию и не позволяет однозначно привязывать аномальные скачкообразные изменения проводимости к перечисленным выше классическим механизмам, предсказывающим подобные скачки.

Среди экспериментальных исследований переключений проводимости, выполненных в работах [7-13] в ПВХ композитных пленках, можно выделить аналоги известных из литературы явлений, в частности скачкообразные гигантские изменения проводимости на вольтамперных характеристиках полимерных пленок типа «мягкого» пробоя [3,5], скачки проводимости под воздействием одноосного давления [6], экспоненциальный рост сопротивления при увеличении толщины образца [4]. Специфика и новизна результатов полученных в работах [7-13] состоит в том, что в отличие от выполненных ранее экспериментов, исследования осуществлялись и на относительно толстых пленках порядка сотен микрон, приготовленных методом полива из растворов композитов ПВХ, что сразу позволило «отсечь» ряд обсуждавшихся в [3] механизмов. Более того, кроме аналогов известных скачкообразных переключений проводимости в экспериментах [7,13] приведены также не описанные ранее (на наш взгляд) явления, связанные с появлением запаздываний и неоднозначности типа гистерезиса на вольтамперных характеристиках, проявляющиеся также в спонтанных переходах в состояние высокой проводимости (СВП) и обратно при неизменном напряжении источника (см. рис. 1 ).

Fig. 1. Спонтанное переключение в состояние высокой проводимости в пленке
сополимера ПВХ-ПАц. ( Spontaneous transition to HCS for PVC- polyacetylene

nanocomposite film)

Как видно из рис.1, ток **I** самопроизвольно и достаточно быстро изменяется в широких пределах. Более подробные измерения [12,13] указывают на спонтанные изменения проводимости в пределах двенадцати порядков величины. В этих условиях о воспроизводимости и однозначности вольтамперной характеристики можно говорить только в отсутствие подобных скачков проводимости. Тем не менее следует отметить, что при увеличении приложенного напряжения переход в СВП неминуемо происходил практически для всех исследованных образцов, т.е. существует диапазон напряжений, в котором аналог «мягкого пробоя» преобладает, причем с ростом напряжения вероятность спонтанного перехода в СВП так же увеличивается. Отметим, что и в этих условиях мягкого пробоя однозначности вольт-амперных характеристик нет [1,2] и образец может остаться в СВП (куда его всегда можно перевести, увеличивая приложенное напряжение) даже при полном снятии напряжения.

Таким образом, в дополнение к двум классическим, обсуждавшимся выше механизмам формирования скачка проводимости необходимо привлечь новый физический механизм «ответственный» за формирования задержки «свитчинга» и появления необратимости вольтамперной характеристики. Из общих соображений привлекаемый механизм должен удовлетворять нескольким требованиям :

- во первых он должен вводить случайную временную задержку в широком диапазоне временных интервалов - от долей секунды до суток и более;

– во вторых он должен обеспечивать обратимость переходов во всем широком диапазоне времен жизни фиксированных состояний проводимости;

– и наконец, главное условие, он должен непосредственно влиять на проводимость материала с тем, чтобы в соответствии с моделью, развитой в работах [1,2], обеспечивать неустойчивость любого из фиксированных состояний относительно возмущений структуры, или иными словами, положительную обратную связь, посредством которой формируется переход в СВП типа перескока, а не плавное «релаксационное переползание» от низких к высоким уровням проводимости и обратно. Поскольку характерные времена сохранения заданных состояний в эксперименте могут иметь значения от единиц секунд до нескольких часов, то очевидно, что механизм формирования таких разных, в том числе и продолжительных времен скорее всего может быть связан только с химическими реакциями. **Для ПВХ композитов химический процесс, обладающий перечисленными выше свойствами известен и называется реакцией дегидрохлорирования, т.е. отделения молекулы HCl от макромолекулы ПВХ с образованием двойной связи (что способствует резкому увеличению проводимости) .** Отметим, что если этот процесс происходит внутри образца ПВХ, то молекуле HCl, отделившейся от макромолекулы полимера, «некуда бежать» и в лучшем случае, она может отодвинуться от «своего места» в макромолекуле на несколько нанометров вследствие наличия в полимере свободного объема, т.е. не плотной упаковки макромолекул полимера. Находясь в непосредственной близости от двойной связи, молекула HCl может вновь войти в структуру макромолекулы, обеспечивая широкий разброс характерных времен релаксации и в то же время обратимость переходов между состояниями проводимости.

Для подтверждения предлагаемого механизма получим численную оценку локального разогрева полимерной пленки толщиной 10 мкм в условиях, близких к экспериментальным [6-13], т.е. при напряжении 1В и токе десять пикоампер (сопротивление пленки $10^{11}$ Ом ). Учтем при этом, что процесс протекания тока, как известно, осуществляется в тонких токоведущих каналах размером порядка 10 мкм. Для получения численной оценки, подсчитаем разогрев вещества в таком токоведущем канале: полная мощность источника $10^{-11}$ Дж/сек, объем канала с линейным размером $10^{-4}$ см получим $10^{-12}$ см$^3$ . Подставляя плотность и теплоемкость ПВХ получаем оценку изменения температуры в секунду из выражения:

- $\Delta T = \Delta Q / m_k C_{ПВХ} = (U \cdot I)/(\rho_{ПВХ} C_{ПВХ} V_к)$                   (1)

где $\Delta T$=изменение температуры за одну секунду , U – напряжение на образце, I – ток через образец, $\rho_{ПВХ}$ =1.4 г/см$^3$ , $C_{ПВХ}$ = 0.09 Дж/Г*К- плотность и теплоемкость ПВХ (справочные данные) и $V_k$ - объем токоведущего канала, в нашем случае $10^{-12}$ см$^3$ . В результате вычислений получаем оценку $\Delta T$=79$^0$ . Таким образом, по порядку величины мы попадаем в диапазон температур начала активного дегидрохлорирования ПВХ (от 70 до 200 $^0$C ). Разумеется, наиболее критическим для данной оценки является размер токоведущего канала, однако, как по литературным данным [1], так и в экспериментальных работах [1-13] , размер токоведущего канала несущественно отличался от приведенной цифры, хотя она, конечно, не является какой-либо фиксированной константой и скорее всего может колебаться в широких пределах в зависимости от используемого материала, геометрии образца и  условий эксперимента. С другой стороны, и напряжения, прикладываемые к образцу, и, соответственно, ток также изменялись в экспериментах в достаточно широких пределах так, что порядок оценки изменения температуры в канале безусловно может попадать в диапазон локального развития реакции дегидрохлорирования. Кроме того отметим, что получена оценка скорости выделения тепла, причем важен именно порядок величины, поскольку очевидно, что вследствие низкой теплопроводности полимерных материалов точное решение уравнения теплопроводности может привести к существенному увеличению $\Delta T$. Таким образом, можно считать полученную оценку «оценкой снизу».  Реально температуры в токоведущем канале могут быть существенно выше, при этом заведомо реализуется реакция дегидрохлорирования, в результате которой молекулы HCl отодвигаются на единицы нанометров ( в рамках свободного объема полимерного материала) . При этом проводимость за счет образовавшихся двойных связей  может  увеличиваться на порядки величины и происходит переключение в СВП. При снятии (или существенном изменении) напряжения с образца  проводимость, связанная, как отмечалось выше с химической структурой токоведущего канала, может  еще некоторое, достаточно продолжительное  время сохраняться. Однако, поскольку молекула H-Cl в толще полимера не имеет возможности далеко отойти от «своего места»  в макромолекуле, происходит обратная реакция присоединения HCl на место двойной связи и перехода в исходное состояние низкой проводимости.

    Как было показано в [12, 13] композитные материалы на основе ПВХ , в частности, подвергнутые частичному дегидрохлорированию в растворе, с получением сополимера ПВХ-ПАц гораздо легче переходили в состояние высокой проводимости, при этом перепад сопротивлений составлял рекордную величину 12 порядков величины. Это также можно рассматривать как подтверждение существенного изменения электропроводности именно в процессе реакции дегидрохлорирования.

    Особенно наглядно предлагаемая модель «химического» теплового переключения проводимости может быть проиллюстрирована с привлечением понятий теории протекания для композитных материалов на основе ПВХ, состоящих из проводящих элементов и изолирующих промежутков[14].  Для простоты рассмотрим решеточную модель среды вблизи порога протекания, т.е. в области концентраций проводящих элементов, близких к образованию бесконечного кластера (БК).  Пусть добавление одного проводящего элемента реализует образование БК, тогда  удаляя этот «последний замыкающий»  проводящий элемент получим, что вся энергия тока будет выделяться в объеме равном одному проводящему элементу, при этом даже очень небольшое тепловыделение способно нагреть изолирующий промежуток ПВХ до температур, соответствующих  протеканию реакции дегидрохлорирования.

**Выводы.**

1. В рамках работ [1,2] показано, что при наличии времени задержки состояние проводимости полимерного композита неустойчиво, т.е. существует положительная обратная связь, подхватывающая случайные возмущения в форме появления свободных зарядов и локального возрастания токов. Это объясняет наблюдаемые в эксперименте «гигантские» скачки проводимости как спонтанные, так и при изменении одноосного давления, освещения, температуры и других внешних воздействиях.

2. Если в среде отсутствуют процессы изменения проводимости, характеризуемые конечными временами задержки, превышающими временное разрешение измерений, то все процессы обратимы, отображаются неизменной вольтамперной характеристикой, фиксированных состояний и спонтанных переходов между этими состояниями нет

3. Предложенный в работе физический механизм реализации временной задержки позволяет полностью снять «аномальность» с явлений, исследованных в работах [6-13] в полимерных композитах на основе ПВХ. Таким образом, наблюдаемые аномалии электропроводности, включая спонтанные и стимулированные переходы в высокопроводящее состояние могут быть описаны в рамках модели, включающей классическую теорию протекания, эффект туннелирования и, согласно предлагаемой модели, тепловое воздействие на макромолекулу ПВХ, неустойчивую относительно нагрева на сотню градусов, с развитием реакции дегидрохлорирования (т.е. отделения молекулы HCl с образованием двойной связи). Развитая модель дает простейшую качественную интерпретацию спонтанных переключений проводимости: — причина: нагрев токоведущего канала, момент переключения — прохождение в достаточной для изменения проводимости реакции дегидрохлорирования (или обратной реакции) в канале проводимости.

В заключение отметим, что область применимости модели может выходить за рамки только ПВХ композитов, поскольку известны аналогичные реакции теплового разложения полимеров с отщеплением фрагментов из макромолекул, в частности, известна реакция дегидрирования с возможным образованием двойных связей. Отметим, что в данной серии экспериментов на ПВХ наблюдается активная перколяция, или иначе, перколяция с самоорганизацией устойчивых связей. Расшифрованное аномальное поведение композитов ПВХ может быть использовано для создания элементов памяти, ключевых элементов, а также мемристоров и моделирования сложных систем с самоорганизацией устойчивых электропроводящих каналов.

Литература